\documentclass[aps,prl,twocolumn,showpacs,superscriptaddress]{revtex4-1}

\usepackage{graphicx}
\usepackage[dvips]{color}

\begin{document}

\title{ Competition between electron and phonon excitations in the scattering of
nitrogen atoms and molecules off tungsten and silver surfaces}

\author{L. Martin-Gondre}
%
%\email{ludovic\_martin@ehu.es}
%
\affiliation{Centro de F\'{\i}sica de Materiales CFM/MPC (CSIC-UPV/EHU), San
Sebasti\'an, Spain}
\affiliation{Donostia International Physics Center DIPC, San Sebasti\'an,
Spain}
\author{M. Alducin}
%
%\email{wapalocm@sq.ehu.es}
%
\affiliation{Centro de F\'{\i}sica de Materiales
CFM/MPC (CSIC-UPV/EHU), San Sebasti\'an, Spain}
\affiliation{Donostia International Physics Center DIPC, San Sebasti\'an,
Spain}
\author{G. A. Bocan}
%
%\email{gbocan@gmail.com}
%
\affiliation{Centro At\'omico Bariloche, CNEA and CNICT, S.C. de Bariloche,
Argentina}
\author{R. D\'{\i}ez Mui\~no}
%
%\email{rdm@ehu.es}
%
\affiliation{Centro de F\'{\i}sica de Materiales CFM/MPC (CSIC-UPV/EHU), San
Sebasti\'an, Spain}
\affiliation{Donostia International Physics Center DIPC, San Sebasti\'an,
Spain}
\author{J. I. Juaristi}
%
%\email{josebainaki.juaristi@ehu.es}
%
\affiliation{Departamento de F\'{\i}sica de Materiales, Facultad de
Qu\'{\i}micas, UPV/EHU, San Sebasti\'an, Spain}
\affiliation{Centro de F\'{\i}sica de Materiales CFM/MPC (CSIC-UPV/EHU), San
Sebasti\'an, Spain}
\affiliation{Donostia International Physics Center DIPC, San Sebasti\'an,
Spain}

\date{March 5, 2012}

\begin{abstract}
We investigate the role played by  electron-hole pair and phonon excitations in
the interaction of reactive gas molecules and atoms with metal surfaces. We
present a theoretical framework that allows us to evaluate within a
full-dimensional dynamics the combined contribution of both excitation mechanisms
while the gas particle-surface interaction is described
by an ab-initio potential energy surface. The model is applied to study energy dissipation
in the scattering of N$_2$ on W(110) and N on Ag(111). Our results show that
phonon excitation is the dominant energy loss channel whereas electron-hole
pair excitations represent a minor contribution. We substantiate that, even
when the energy dissipated is quantitatively significant, important aspects of
the scattering dynamics are well captured by the adiabatic approximation.
\end{abstract}

\pacs{82.65.+r,34.35.+a,34.50.-s,68.49.Df,82.20.Kh,82.20.Rp}

\maketitle

In the last years, with the development of ab-initio calculations based on
density functional theory (DFT), unprecedented accuracy has been achieved in
describing the interaction of reactive thermal and hyperthermal gas molecules
and atoms with metal surfaces. In most advanced simulations, molecular dynamics
are performed in ab-initio (DFT) ground state multidimensional potential energy
surfaces (PES)~\cite{gross03}. This refined scheme relies on the validity of
the adiabatic approximation. By adiabatic, we refer to a process which
neglects both electronic
excitations and lattice phonons excitations. Nevertheless, the role played
by these nonadiabatic effects is now under close scrutiny. When the
experimental results and the adiabatic results are at variance, quite often
controversy arises about up to what extent the differences are due to the
neglect of energy loss channels~\cite{luntz05}, to the limitations of a reduced
dimensionality approximation~\cite{diaz06}, or to the inherent limitations of
DFT~\cite{bocan08,disc09,geetha11}.

The increasing number of gas-surface experiments reporting evidence of
electronic excitations~\cite{gergen01,white05}, has motivated the
development of different models to include this dissipation
channel~\cite{trail03,juaristi08,shenvi09,monturet10}. Among them, the
local density friction coefficient approximation
(LDFA)~\cite{juaristi08} offers a very good compromise between the
accuracy of the results and the simplicity of its
implementation~\cite{tremblay10}. Different approaches have been also
developed to treat the energy exchange with the
lattice~\cite{tully80,kimman86,tenner91,manson91}. Semiclassical
approximations of the phonon excitations, in which the gas-surface
interaction is limited to simplified model potentials~\cite{manson91},
have been very successful in understanding the scattering of
nonreactive rare gas atoms with surfaces~\cite{hayes07}. However, they
are less accurate when applied to reactive species~\cite{ambaye06}. In
the latter case, the multidimensional PES is typically so corrugated
and intricate that a more realistic treatment of the interaction is
required. The generalized Langevin oscillator model
(GLO)~\cite{tully80} shows to be a sound alternative in this
respect~\cite{busnengo04}.

Still, the challenge in gas-surface dynamics is to provide a
theoretical framework that, keeping the accuracy of a multidimensional
ab-initio PES for the gas-metal interaction, incorporates into the
dynamics energy exchange with {\it both} lattice vibrations and
electronic excitations~\cite{krsc08,hasc09}. In this Letter we
accomplish this objective by combining the GLO for phonon excitations
and the LDFA for electronic excitations. The inclusion of both effects
will allow us to address such fundamental questions as (i) what is the
relative importance of phonon and electron-hole ({\em e-h}) pair
excitations as energy dissipation channels, in particular, (ii) is
there any coupling between them or is their contribution just
additive, and (iii) to what extent does the adiabatic calculation
capture the basic physics of the dynamics and provide accurate
results.

To answer these questions, we have selected two systems for which accurate
energy loss measurements exist: the rotationally inelastic scattering of
N$_{2}$ on W(110)~\cite{hanisco93} and the scattering of hyperthermal N on
Ag(111)~\cite{ueta09}. Molecular and atomic nitrogen are relatively
heavy
projectiles for which energy exchange to the
lattice can be important. The use of N also implies that electronic excitations
are analyzed under the most favored conditions of open-shell reactive species.

The way in which we incorporate electronic and phonon excitations into the
multidimensional classical trajectory simulations is as
follows~\cite{[See Supplemental Material at ] [ for a detailed
description.]supp}. As in the GLO, the motion of surface atoms is represented
by a three-dimensional (3D) harmonic oscillator. The latter is coupled to a 3D
ghost oscillator that is subjected to frictional and random forces accounting
for energy dissipation and thermal fluctuations~\cite{busnengo04}. The effect
of electronic excitations is added by introducing for each of the impinging
gas atoms a separate friction force proportional to its
velocity. The electronic friction coefficient is calculated at each point of
the trajectory as that of the atom moving in a homogeneous free electron gas
with electronic density equal to that of the surface at this
point~\cite{juaristi08}. Very recently, it has been shown that
this model constitutes an efficient and sufficiently accurate tool to incorporate
electronic excitations within multidimensional molecular dynamics~\cite{Fuchsel2011}.
\begin{figure}
%\suppressfloats
\resizebox{!}{0.45\textwidth}{
\includegraphics*{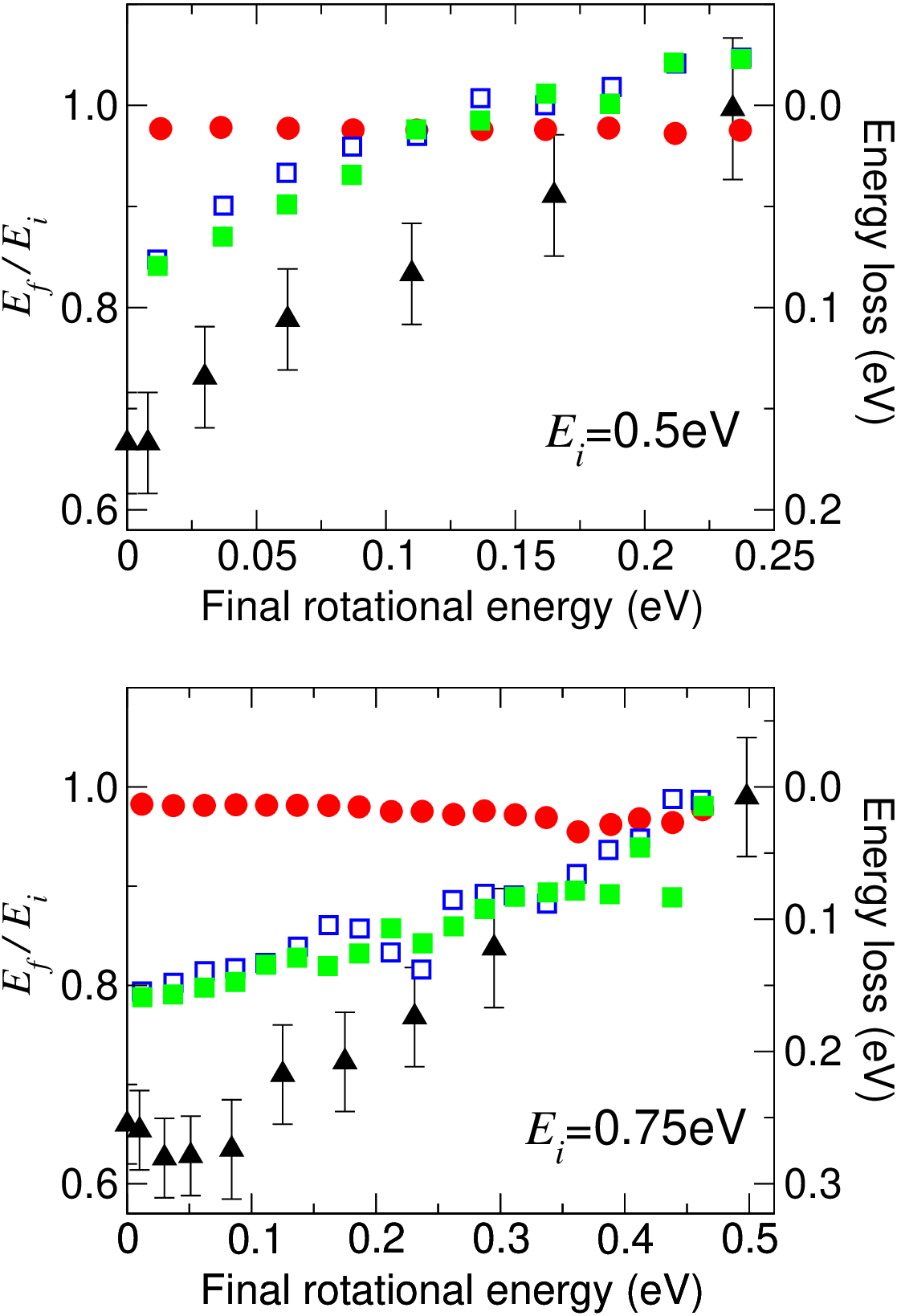}}
\caption{(Color online) Total final energy normalized to the incident
energy (left ordinate) and energy loss (right ordinate)
 of N$_2$ scattered off W(110)
vs the exit rotational energy. Results for normal incidence and
detection angles and two incidence energies $E_i$. In triangles, the
experimental data of~\cite{hanisco93}. Our simulations are represented
by red filled circles (LDFA), blue open squares (GLO), and green
filled squares (full nonadiabatic calculation). $T_s=1200$~K.
\label{f:s1} }
\end{figure}

When a rotationally cold N$_2$ beam is scattered off W(110), the molecules lose
around 30\% of their incidence energy for low exit rotational states, but the
energy loss is lower for those scattered at high exit rotational
states~\cite{hanisco93}. To understand these findings, we have performed
classical molecular dynamics simulations with different approximations: (i)
using the adiabatic approximation, i.e. neglecting {\em e-h}
pair and phonon excitations, (ii) including only {\em e-h} pair excitations
(LDFA), (iii) including only energy exchange with the lattice (GLO) and (iv)
including both energy dissipation channels (full nonadiabatic calculation with
respect to both electrons and phonons). In all cases we use the 6D ab-initio
N$_2$/W(110) PES of~\cite{aldiprl06,alducin06}, where the PW91
exchange correlation functional was used~\cite{perdew92}. A minimum of 30000
trajectories is calculated using a conventional Monte Carlo sampling of all
possible initial conditions. In (iii) and (iv), the parallel and perpendicular
surface oscillator frequencies are 19 and 16~meV respectively~\cite{balden96},
and the friction coefficients of the ghost oscillators are obtained from the
Debye frequency as proposed in~\cite{tully80}.

In Fig.~\ref{f:s1} we compare the results of our simulations with the
experiments. The figure shows the fraction of energy retained and the
energy loss as a function of the exit rotational energy for the
scattered molecules, at normal incidence and detection angles. When
only {\em e-h} pair excitations are included, contrarily to the
experimental observations, the energy loss, which is marginal, does
not depend on the exit rotational energy. It only depends on the total
energy no matter how it is distributed among the different degrees of
freedom. The inclusion of phonon excitations changes the picture
completely. We recover the experimental observation that more energy
is lost at low exit rotational states. Note also that when phonon
excitations are included the results with and without {\em e-h} pair
excitations are indistinguishable. The differences are, within the
statistical errors, of the order of the contribution of {\em e-h} pair
excitations alone. This reflects the predominant role of phonon
excitations in this kind of experiment.

The importance of including energy exchange with the lattice can be
rationalized by noticing that at normal incidence and detection,
though corrugation and anisotropy of the PES may complicate the
picture, backscattering conditions must prevail. As a consequence,
large momentum transfer takes place from the projectile to the lattice
in the direction normal to the surface. This implies a comparatively
larger probability for translational energy transfer. As a result,
molecules that are rotationally excited at the expense of their
translational energy in the scattering with the surface, are more
inefficient transferring energy to the lattice. In
Ref.~\cite{kimman86}, a simplified kinematic model has been proposed
to illustrate this effect. Note that at the highest rotational states,
within the experimental error bars, it cannot be decided whether the
molecules overcome minor energy losses or energy gains. In fact, for
E$_i$=0.5 eV our calculations give a small energy gain consistent with
the experimental error bars. In this case, the efficient conversion of
translational energy into rotational energy implies a big reduction of
the former, and, for the high temperatures under consideration
($T_s=1200$~K), energy transfer from the lattice to the projectile is
favored. The {\em e-h} pair excitation mechanism cannot give rise to
this kind of behavior due to the large mismatch between the projectile
and the electron masses.

\begin{figure}
\suppressfloats \resizebox{!}{0.25\textwidth}{
\includegraphics*{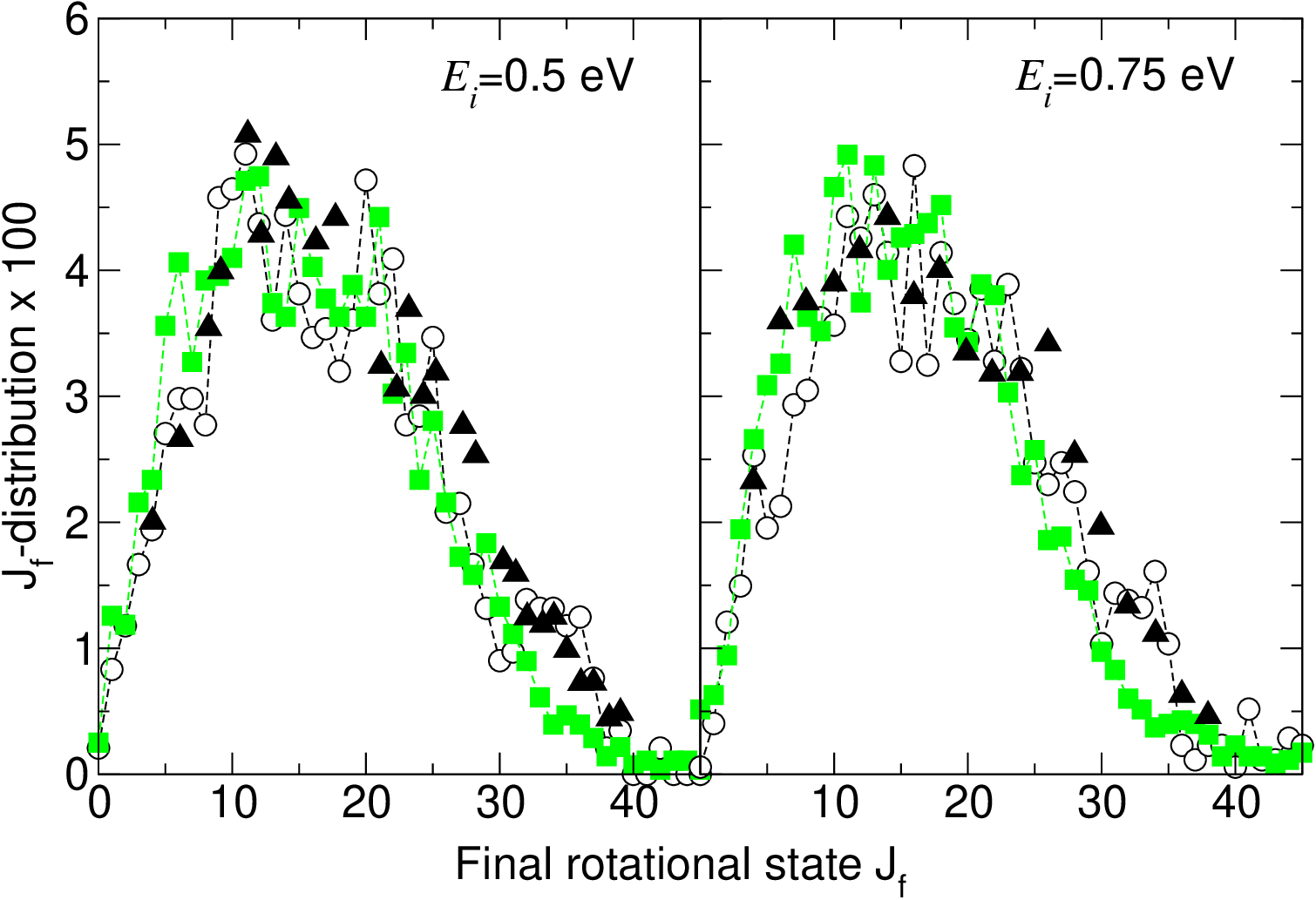}}
\caption{(Color online) Final rotational state population
distributions for N$_2$ scattered from W(110) under the same
experimental conditions of Fig.~\ref{f:s1}. The experimental data
of~\cite{hanisco93} (filled triangles) are compared with the
adiabatic~\cite{geetha11} (open circles) and the full nonadiabatic
(green filled squares) calculations.\label{f:s2}}
\end{figure}
Interestingly, the measured rotational state population distributions
of the scattered N$_2$ are already well reproduced within the
adiabatic approximation, as shown in~\cite{geetha11}. In
Fig.~\ref{f:s2} we show that including energy exchange with lattice
and {\em e-h} pair excitations the width and shape of the
distributions are not altered significantly. In other words, the
conversion from translational to rotational energy in this kind of
experiment is basically an adiabatic process, and the result of the
scattering on a highly anisotropic and corrugated six-dimensional PES.
Hence, even when the molecules lose a significant amount of energy,
the adiabatic approximation is still valid to describe different
aspects of the molecule-surface interaction, provided an accurate
full-dimensional PES is used. The picture that emerges from our
results is the following: in the scattering of the molecules with the
anisotropic and corrugated PES, energy is transferred adiabatically
from the translational to the rotational degrees of motion. Since
translational energy is more efficiently transferred to the lattice
than rotational energy, molecules that remain in the lowest rotational
states (with higher translational energy) are the ones that lose more
energy.

\begin{figure}
\suppressfloats \resizebox{!}{0.25\textwidth}{
\includegraphics*{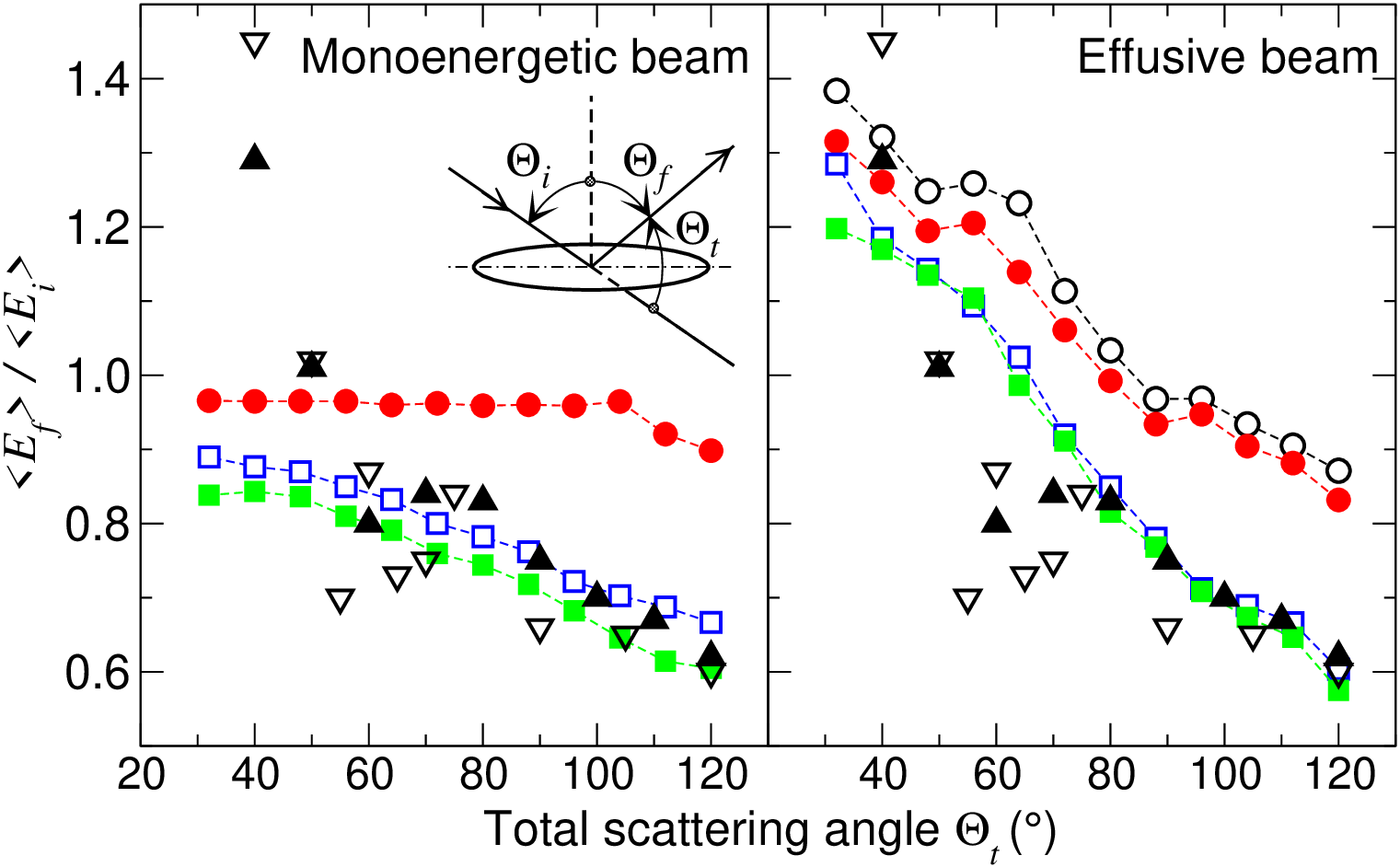}}
\caption{(Color online) Ratio of final to initial average energy vs
the total scattering angle [$\Theta_ t =180^{\circ} - (\Theta_i +
\Theta_f)$] for N/Ag(111) at $T_s=500$~K and $\Theta_ i=60^{\circ}$.
The experimental data of~\cite{ueta09} (open and filled triangles
correspond to different runs of the same experiment) are compared with
our simulations for a monoenergetic (left panel) and for an effusive
beam (right panel): adiabatic results (open circles), LDFA (red filled
circles), GLO (blue open squares), and full nonadiabatic (green filled
squares). \label{f:s3}}
\end{figure}

Next, we use our model to analyze the energy loss observed in the
scattering of N atoms with energies of some eVs on the Ag(111)
surface~\cite{ueta09}. These measurements represent an excellent
benchmark to explore the accuracy of our model for atoms and higher
projectile energies, for which {\em e-h} pair excitations are expected
to be more relevant. Figure ~\ref{f:s3} reproduces the experimental
results for an effusive beam of N scattered from Ag(111) at $T_s=
500$~K and an incidence angle $\Theta_i=60^{\circ}$. The beam has an
average energy of 4.3~eV and a FWHM of $\sim 5.0$ eV. The figure
depicts the ratio between the average final $<E_f>$ and initial
$<E_i>$ energies as a function of the in-plane scattering angle
$\Theta_t$ (see inset). The experiments show a decrease of the final
average energy as the scattering angle increases. A remarkable feature
is the dramatic increase of the energy ratio at $\Theta_t <
60^{\circ}$, i.e., for grazing outgoing angles. In fact, the final
average energy is larger than the initial one in this angular range.

We have performed classical dynamics calculations, using the ab-initio
3D N/Ag(111) PES of~\cite{bocan10}. The results of our different
simulations for a monoenergetic beam with $E_i=$~4.3~eV and
$\Theta_i=60^{\circ}$ are represented in the left panel of
Fig.~\ref{f:s3}. For each kind of simulation, we calculate 300000
trajectories to assure good statistics. The parallel and perpendicular
surface oscillator frequencies used in the GLO and full nonadiabatic
calculations are 14 and 9 meV respectively~\cite{ponjee03}. The
adiabatic results that correspond to $<E_f>/<E_ i>=1$ are not shown in
the figure. Despite the large incident energy, we observe that
\textit{e-h} pair excitations (LDFA calculations) produce marginal
energy losses. In addition, the electronic energy loss is roughly
independent of the scattering angle. When energy exchange with the
lattice is included (GLO and full nonadiabatic calculations), the
experimental results for large scattering angles are well reproduced.
Again, differences between the results of these two models are of the
order of the contribution of {\rm e-h} pair excitations alone.
However, none of the energy dissipation channels is able to explain
the energy ratios larger than 1 that are measured for small scattering
angles. We find that this behavior is indeed due to the effusive beam
itself.

The results of our calculations mimicking the experimental effusive beam are
shown in the right panel of Fig.~\ref{f:s3}. In this case, we use 600000
trajectories. For large scattering angles ($\Theta_t \ge 80^{\circ}$), the
results do not significantly differ from those obtained with the monoenergetic
beam. However, at small scattering angles, all the effusive beam simulations
give final to initial energy ratios greater than one. Quantitative agreement
with the experiments is obtained only when energy exchange with the lattice is
allowed, no matter whether \textit{e-h} pair excitations are included or not.
The small discrepancy observed at $\Theta_t \sim 60^{\circ}$ might be due to
the composition of the N experimental beam containing not only the ground state
N$(^{4}S)$ but also electronically excited states N$(^{2}D)$, N$(^{2}P)$,
assumed to be mainly reflected around the specular position ($\Theta_t =
60^{\circ}$)~\cite{ueta09}.

A capture mechanism has been proposed to explain the small-$\Theta_t $
behavior~\cite{ueta09}. This capture mechanism assumes that the lower energy
atoms with grazing exit trajectories are trapped close to the Ag surface,
resulting in the increase of the average final energy. An analysis of our
simulations shows that the small-$\Theta_t $ behavior is due to the fact
that the higher energy atoms ($E_i \ge $ 7 eV) of the effusive beam
are preferentially reflected at low scattering angles. As a result, the
average energy of the atoms scattered at small (large) $\Theta_t$ is
higher (lower) than the average energy of the incident beam. We stress
that this is a pure trajectory effect that is not related to the presence
or not of inelastic channels: our
adiabatic simulations for the effusive beam already reproduce the correct
dependence of $<E_f>/<E_ i>$ on $\Theta_t$. This clearly
indicates that a realistic PES is essential in order to reproduce in
quantitative terms this nontrivial dynamic effect.

In summary, we have developed a theoretical framework that allows the
inclusion of both phonon and {\em e-h} pair excitations in a
full-dimensional dynamics, keeping the accuracy of an ab-initio PES to
treat the particle-surface interaction. Though the model is not
applicable to systems in which nonadiabatic effects are due to
crossing of potential energy surfaces with possible charge transfer,
it can provide a proper description for a great variety of systems.
Concerning the three questions posed above, our analysis shows that
for two representative systems: (i) phonon excitation is the main
energy dissipation channel even for the hyperthermal N beams, (ii) the
contribution of phonon and {\em e-h} pair excitations to the total
energy loss is additive, and, finally, (iii) even when energy loss
processes are significant, important aspects of the scattering
dynamics are already captured by the adiabatic calculation: the
conversion of translational to rotational energy upon scattering with
the surface in the N$_2$/W(110) case and the decrease in the number of
low energy atoms that are scattered off the surface with small grazing
exit angles in the N/Ag(111) case. The correct description of these
features requires a full-dimensional dynamics calculation on top of an
accurate PES. In this respect, the theoretical framework proposed here
is an excellent choice to efficiently include all the basic
ingredients in gas-surface simulations.

\begin{acknowledgments}
This work has been supported in part by the Basque Departamento de
Educaci\'{o}n, Universidades e Investigaci\'{o}n, the University of the Basque
Country UPV/EHU (Grant No. IT-366-07) and the Spanish Ministerio de Ciencia e
Innovaci\'on (Grant No. FIS2010-19609-C02-02). Computational resources were
provided by the DIPC computing center.
\end{acknowledgments}

%\bibliography{references-shorten}

\end{document}